\documentclass[11pt]{article}
\usepackage{a4wide}                      
\usepackage{epsf}                        
\usepackage{latexsym}                    
\usepackage{amsfonts}                    
\usepackage{amssymb}
\usepackage{amsmath,amsthm}
\usepackage{graphicx}

\newcommand{\sect}[1]{\setcounter{equation}{0}\section{#1}}
\renewcommand{\theequation}{\arabic{section}.\arabic{equation}}

\def\be{\begin{equation}}
\def\ee{\end{equation}}
\def\bea{\begin{eqnarray}}
\def\eea{\end{eqnarray}}
\def\nn{\nonumber \\}
\def\nnw{\nonumber \\ [.2cm]}

\def\hsp#1{\hspace*{#1}}

\def\part{\partial}
\def\tfrac#1#2{{\textstyle{\frac{#1}{#2}}}}
\def\half{\tfrac{1}{2}}

\def\hF{\hat F}

\def\hs{\hat s}
\def\bh{\bar h}
\def\bK{\bar K}
\def\bR{\bar R}

\def\tg{{\tilde g}}

\def\tR{{\tilde R}}
\def\tLambda{{\tilde \Lambda}}

\def\d{{\mbox{d}}}
\def\e{{\mbox{e}}}
\def\sqrthg{\sqrt{|\hat g|}}
\def\sqrtg{\sqrt{|g|}}

\def\S{\ensuremath{\mathbb{S}}}

\def\hmu{\hat{\mu}}

\def\hphi{{\hat{\phi}}}

\def\mn{{\mu\nu}}

\def\makeatletter{\catcode`\@=11}
\makeatletter
\def\mathbox#1{\hbox{$\m@th#1$}}%
\def\math@ccstyles#1#2#3#4#5#6#7{{\leavevmode
      \setbox0\mathbox{#6#7}%
      \setbox2\mathbox{#4#5}%
      \dimen@ #3%
      \baselineskip\z@\lineskiplimit#1\lineskip\z@
      \vbox{\ialign{##\crcr
             \hfil \kern #2\box2 \hfil\crcr
             \noalign{\kern\dimen@}%
             \hfil\box0\hfil\crcr}}}}
\def\mathaccstyles{\math@ccstyles\maxdimen}
\def\maththroughstyles{\math@ccstyles{-\maxdimen}}
\def\unity%
 {\maththroughstyles{.45\ht0}\z@\displaystyle {\mathchar"006C}\displaystyle 1}
\begin{document}
\rightline{UUITP-20/11} \rightline{UG-FT-288/11}
\rightline{CAFPE-158/11}
\rightline{\today}
\vspace{1truecm}

\centerline{\huge \bf Einstein Branes }
\vspace{1.3truecm}

\centerline{
    {\large \bf Wissam Chemissany${}^{a}$, } \
    {\large \bf Bert Janssen${}^{b}$} \
    {\bf and}  \
    {\large \bf Thomas Van Riet${}^{c}$}
}

\vspace{.4cm}
\centerline{{\it ${}^a$  Afdeling Theoretische Fysica, Katholieke Universiteit Leuven }}
\centerline{{\it Celestijnenlaan 200D bus 2415, 3001 Heverlee, Belgium}}
\centerline{{\tt wissam@itf.fys.kuleuven.be} }

\vspace{.4cm}
\centerline{{\it ${}^b$ Departamento de F\'{\i}sica Te\'orica y del Cosmos and}}
\centerline{{\it Centro Andaluz de F\'{\i}sica de Part\'{\i}culas Elementales}}
\centerline{{\it Universidad de Granada, 18071 Granada, Spain}}
\centerline{{\tt bjanssen@ugr.es} }

\vspace{.4cm}
\centerline{{\it ${}^c$ Institutionen f\"or fysik och astronomi}}
\centerline{{\it Uppsala Universitet, Box 803, SE-751 08 Uppsala, Sweden}}
\centerline{{\tt thomas.vanriet@fysast.uu.se} }

\vspace{2truecm}

\centerline{\bf ABSTRACT}
\vspace{.5truecm}

\noindent We generalise the standard, flat $p$-brane solutions sourced by a
dilaton and a form field, by taking the worldvolume to be a curved
Einstein space, such as (anti-)de Sitter space. Our method is based
on reducing the $p$-branes to domain walls and then allowing these
domain walls to be curved.  For de Sitter worldvolumes this extends
some recently constructed warped de Sitter non-compactifications. We
restrict our analysis to solutions that possess scaling behavior and
demonstrate that these scaling solutions are near-horizon limits of
a more general solution. Finally, our framework can equally be used
for spacelike branes and the uplift of the domain wall/cosmology
correspondence becomes in this context a more general
timelike/spacelike brane correspondence.

\newpage
\sect{Introduction}

The r\^ole of $p$-branes in string theory and supergravity can
hardly be overestimated. They appear as fundamental objects in
string theory and as (solitonic) solutions of the equations of
motion of supergravity and are crucial for the understanding of many
aspects of string theory, such as gauge/gravity duals, string
phenomenology, string cosmology and non-perturbative effects, to
mention a few.

$p$-branes typically arise as solutions of the equations of motion
of supergravity-like actions that describe $D$-dimensional Einstein
gravity coupled to a dilaton and form-fields, \be S_D = \int \d^D x
\sqrthg
      \left[ \hat R \ - \ \frac{1}{2} (\partial \hphi)^2
          - \frac{1}{2(p+2)!} \,  \e^{b\hphi}
                  \hat F_{\hmu_1 \dots \hmu_{p+2}}\hat F^{\hmu_1 \dots \hmu_{p+2}} \right].
\label{action}
\ee
Most of the fundamental $p$-brane solutions in ten and eleven dimensions are known since the
1990's \cite{Dabh}-\cite{Polch} and the general solution of the type
\bea
&& \d\hat s^2 \, = \, H^{2A}(r) \, \eta_{ij} \,\d x^i \d x^j \,
           +  \, H^{2B}(r) \Bigl( \d r^2 + r^2 \d\Omega_n^2\Bigr),\nnw
&& e^{-2\hphi} \, = \, H^C(r), \hspace{2cm}
\hF_{i_1 \dots i_{p+1}r} \, = \, \partial_r H^E(r) \,\varepsilon_{i_1 \dots i_{p+1}},
\label{Ansatz}
\eea
can be found in, for example, \cite{LPSS,Bergsh}, with the constants $A$, $B$, $C$ and $E$ and the
amount of preserved supersymmetry depending on the parameters $D$, $p$ and $b$ in the action
(\ref{action}).

There are several known generalisations of the above (flat)
$p$-brane Ansatz. In \cite{BP}-\cite{FoF} it was shown that the
Ansatz (\ref{Ansatz}) can be easily generalised to include a curved
worldvolume metric $\tg_{ij} = \tg_{ij}(x)$, without changing the
constants $A$, $B$, $C$ and $E$, as long as the worldvolume geometry
is Ricci-flat, $\tR_{ij}= 0$. In that case, the amount of
supersymmetry of the Ricci-flat solution reduces to the number of
Killing spinors of $\tg_{ij}$.  On the other hand, a vast list of
domain wall solutions is known (either exactly or numerically),
whose worldvolume has constant curvature, or at least has an
Einstein geometry, $\tR_{ij}= \tLambda \, \tg_{ij}$
\cite{KSS}-\cite{ST}. Recall that domain walls are special kind of
$p$-branes, defined by having co-dimension one only and that they are
magnetically charged with respect to a 0-form ``field strength''.
This means we can find such solutions from in theories with gravity coupled to a
scalar with some scalar potential:
\begin{equation}
S_D =  \int \d^D x \sqrthg
      \left[ \hat R \ - \frac{1}{2} (\partial \hphi)^2
          - V(\hphi) \right] .
\end{equation}

Curiously enough, to our knowledge, there are very few known
$p$-brane solutions for general $p$ ({\it i.e.} not domain walls)
that have an Einstein geometry in their worldvolume, and even so,
they are sometimes not interpreted as such. Yet the construction of
these solutions would be very interesting: $p$-branes with a
positive worldvolume curvature could yield an alternative way to
obtain de Sitter-like solutions in supergravity in the sense of
Randall-Sundrum \cite{Randall:1999vf} (see also
\cite{Fonseca:2011ep}). But also in the context of compact extra
dimensions curved brane solutions are relevant. A generic flux
compactification down to (anti-)de Sitter space involves brane
sources that fill the lower-dimensional non-compact space and wrap
some submanifold in the internal space (or are simply pointlike in
the internal space). This implies that the brane's worldvolume is
necessarily curved. In most cases these solutions are only
understood in the limit that the branes are smeared over the
internal space. The full backreacted solutions turn out subtle and
might not always exist \cite{Douglas:2010rt}-\cite{Blaback:2011nz}. 
Finally, $p$-branes with an AdS geometry in their worldvolume might
give rise to a new class of supersymmetric solutions, as domain
walls with AdS curvature have been observed to be supersymmetric in
specific cases.

It is well known that there is a one to one correspondence between
$D$-dimensional $p$-branes and domain walls in $(p+2)$
dimensions, by reducing the solutions  (\ref{Ansatz}) over the
angular part $\d\Omega_n^2$ of the transverse space, where after reduction the
radial coordinate $r$ corresponds to the direction transverse to the
domain wall. This suggests a obvious way to find $p$-branes with Einstein geometry, 
by lifting up curved domain walls to $D = p+n+2$ dimensions.

Concretely, our strategy to construct $p$-brane solutions with Einstein worldvolume will be
that of mapping the most general two-block Ansatz for $D$-dimensional curved branes
\bea
&& \d\hat s^2 \, = \, \e^{2A(r)} \, \tg_{ij} (x)\,\d x^i \d x^j \,
           +  \, \e^{2B(r)} \Bigl( \d r^2 + r^2 \d\Sigma_n^2\Bigr),
\label{Ansatzcurved}
\eea
onto a $(p+2)$-dimensional domain wall Ansatz
\bea
&& \d s^2 \, = \, a^2(z) \tg_{ij}(x) \,\d x^i \d x^j \, + \, f^2(z) \d z^2 , 
\label{DWAnsatz} \eea by reducing over the angular part $\d
\Sigma_n^2 = \bh_{ab}(\theta)\d\theta^a \d\theta^b$ of the
transverse space and than look for solutions whose worldvolume
metric $\tg_{ij}$ is Einstein. The angular metric $\bh_{ab}(\theta)$
is allowed to have a positive, negative of zero Einstein curvature,
\be \bR_{ab} = (n-1)\bK  \bh_{ab}, \hsp{2cm} \bK = 0, \pm 1.
\label{maxsymmh} \ee For convenience we prefer to work with the
dual formulation of the gauge field, where the $p$-brane is
\emph{magnetically} charged under a $n$-form fields strength, with $n=D-p-2$.


\sect{Reduction to a domain wall problem}
\label{reduction}
%

As explained in the introduction, our strategy to obtain curved
$p$-brane solutions of the form (\ref{Ansatzcurved}) will consist of
reducing the problem to that of finding domain wall solutions with
curved worldvolume.  We therefore consider the following Ansatz for
the metric, dilaton and gauge field\footnote{Our conventions are such that
$\bar \varepsilon^{1\dots n} = 1$ and $\bar\varepsilon_{a_1\dots a_n} =
\bh_{a_1b_1} \dots \bh_{a_n b_n}\bar\varepsilon^{b_1\dots b_n}$.},
\bea && \d\hat s^2 \ = \
\e^{2\alpha\chi}\, g_\mn \, \d x^\mu\d x^\nu
          \ + \ \e^{2\beta\chi}\, \bar h_{ab}\, \d\theta^a  \d\theta^b,\nnw
&& \hat \phi = \phi (x),   \hsp{2cm}
\hat  F_{a_1 \dots a_n} \ = \ \frac{1}{\sqrt{|\bar h|}} \, Q\, \bar\varepsilon_{a_1\dots a_n},
\label{Ansatz2}
\eea
where  the Greek indices $\mu, \nu$ run from 0 to $p+2$, while the Latin indices $a,b$ run from
1 to $n$. The functions $g_\mn(x)$, $\chi(x)$ and $\phi(x)$ depend
on the external coordinates $x^\mu$ and the angular metric $\bh_{ab}(\theta)$ satisfies the
property  (\ref{maxsymmh}). The $(p+2)$-dimensional metric $g_\mn$ contains both the
$(p+1)$-dimensional
worldvolume and the radial coordinate $r$ of the transverse space and should be identified with
the domain wall spacetime (\ref{DWAnsatz}), while $\chi$ is a breathing mode. The metric in the form
(\ref{Ansatz2}) can be obtained from (\ref{Ansatzcurved}) through the identifications
 \be
\e^{\beta\chi(z)} \, = \, \e^{B(r)} r, \hspace{2cm}
\e^{\alpha\chi(z)} \, a(z)\, = \,  \e^{A(r)},
\label{redrel}
\ee
together with the coordinate transformation
\be
\e^{\alpha\chi(z)}\, f(z) \, \d z \ = \ \e^{B(r)} \, \d r.
\ee


Substituting the Ansatz (\ref{Ansatz2}) in the action (\ref{action}), we find that the latter
reduces to the action of $(p+2)$-dimensional gravity coupled to two scalars $\phi$ and $\chi$
in a double-exponential potential,
\be
S_{p+2}  =  \frac{1}{\kappa} \int\d^{p+2} x \sqrtg \Bigl[ R \ - \ \half(\partial\phi)^2
      \ - \ \half(\partial\chi)^2
      \ - \ \half \e^{b\phi + c\chi} Q^2
      \ + \ n(n-1) e^{d\chi}\bK\Bigr],
\ee
where we have imposed the conditions
\be
p\alpha = -n\beta, \hspace{2cm} \alpha = \sqrt{\frac{n}{2p(p+n)}},
\ee
in order to be able to write the action in Einstein frame and to canonically normalise $\chi$,
respectively. Hence $\beta$ and the parameters $c$ and $d$ in the potential are given by
\be
\beta = -\sqrt{\frac{p}{2n(p+n)}},  \hspace{1cm}
c= (p+1) \sqrt{\frac{2n}{p(p+n)}},  \hspace{1cm}
d= \sqrt{\frac{2(p+n)}{pn}}.
\label{values2}
\ee

The curved domain wall solutions we will be interested in are of the type 
\bea && \d s^2 \, =
\, a^2(z) \tg_{ij}(x) \,\d x^i \d x^j \, + \, f^2(z) \d z^2 , \nnw
&& \phi = \phi(z), \hspace{2cm}  \chi = \chi(z), \label{DWAnsatz2}
\eea where the worldvolume metric $\tg_{ij}$ satisfies the Einstein
condition
 \be
\tR_{ij} = p\,\tLambda \, \tg_{ij}. 
\label{4dimcurvature}
\ee
 The equations of motion, for this Ansatz, can be written as
\bea
&& f^2 a^{-2} \tLambda \ + \  \frac{a''}{a}
            \ - \ \frac{f'a'}{fa}
            \ + \ \frac{1}{2p} (\phi')^2
            \ + \ \frac{1}{2p} (\chi')^2 \ = \ 0, \nnw
&& p(p+1) f^2 a^{-2} \tLambda  \ - \ p(p+1) \Bigl( \frac{a'}{a}\Bigr)^2
            \ + \ \frac{1}{2} (\phi')^2
            \ + \ \frac{1}{2} (\chi')^2 \nn
&& \hsp{3cm}
\ -\ \half f^2 \e^{b\phi+c\chi} Q^2 \ + \ n(n-1) f^2 \e^{d\chi}  \bK \ = \ 0,\nnw
&& \phi'' \ + \ (p+1) \, \frac{a'}{a}\, \phi'
          \ - \ \frac{f'}{f}\, \phi'
          \ - \ \half b\, \e^{b\phi + c\chi} f^2 Q^2 \ = \ 0,
\label{EOM} \\[.2cm]
&& \chi'' \ + \ (p+1) \, \frac{a'}{a}\, \chi'
          \ - \ \frac{f'}{f}\, \chi'
          \ - \ \half c\, \e^{b\phi + c\chi} f^2 Q^2
          \ + \ d\, n(n-1) \e^{d\chi} f^2 \bK\ = \ 0,\nonumber
\eea
where a prime denotes differentiation with respect to $z$. A few comments are in order.
The Einbein $f(z)$ is not a dynamical degree of freedom, but a gauge choice, as different choices
of $f(z)$ correspond to different parametrisations of the transverse direction. 
In the forthcoming sections we will usually take $f(z)=1$. This coordinate reparametrisation 
freedom also implies a degeneracy in the above equations of motion.
One can check that only three out of four differential equations are independent, as, for example,
conservation of energy together with the last three yield the first equation. The last three
equations are therefore sufficient to determine the dynamical degrees of freedom $a(z)$, $\phi(z)$
and $\chi(z)$.

Obviously these equations should be able not only to describe curved $p$-brane solutions, but also 
to recover the known standard (flat) $p$-branes (\ref{Ansatz}). Indeed, it is easy to show that for 
example the M2-brane of 11-dimensional
supergravity ($p=2$, $n=7$, $\tLambda= 0$, $\bK= 1$) corresponds to the following solution:
\bea
&& a(r) \, = \, (r^6 + R_0^6)^{1/4}\, r^2, \hsp{1.8cm}
\chi(r) \, = \, \frac{1}{18\sqrt{7}}\, \ln (r^6 + R_0^6), \hsp{1cm}
 \nnw
&& f(r)\, = \, (r^6 + R_0^6)^{3/4}\, r^{-2},  \hsp{1.5cm}
Q \, = \, \pm 6\,R_0^6.
\eea


\sect{Curved branes without flux and the AJS domain wall }
\label{Q=0}

To our knowledge, the first domain wall solution with an Einstein geometry in the worldvolume
was given in \cite{AJS}, which we will refer to as the AJS domain wall. Though the original
derivation was done in five dimensions ($p=3$),
the generalisation to arbitrary $p$ is straightforward. The domain wall appears as a solution
of the Einstein-Hilbert action coupled to a single scalar in an exponential potential,
\be
S_{p+2}  =  \int\d^{p+2} x \sqrtg \Bigl[ R \ - \ \half(\partial\chi)^2
      \ - \ \e^{d\chi}\Lambda \Bigr],
\ee for general scalar coupling $d$ and $(p+2)$-dimensional
cosmological constant $\Lambda$. The AJS domain wall is then given
by\footnote{Using a simple shift of the coordinate $z$, one can write
the scale factor as $a(z)=\lambda z$. This we will occasionally do
in this paper without mentioning. The reason to keep the constant
term is to make a comparison with \cite{AJS}.} \bea \d s^2 \ = \
\Bigl[1 + d\sqrt{\tfrac{-\Lambda}{2p}} \, z\Bigr]^2 \tg_{ij}\, \d
x^i \d x^j
               \ +\ \d z^2,
\hspace{1.5cm}
\e^\chi \ =  \ \Bigl[1 + d\sqrt{\tfrac{-\Lambda}{2p}} \, z\Bigr]^{-\frac{2}{d}},
\label{AJS5}
\eea
where the worldvolume metric $\tg_{ij}$ satisfies the Einstein condition (\ref{4dimcurvature}) and
where the worldvolume curvature $\tLambda$ is determined by the scalar coupling $d$ and the
bulk cosmological constant $\Lambda$,
\be
 \tLambda \ = \ \frac{2 \ - \ p\,d^2}{2p^2} \,  \Lambda.
\ee
The minus sign under the square root in (\ref{AJS5}) allows only negative values for $\Lambda$
and hence we see that we can have de Sitter domain walls for $d^2 > 2/p$ and AdS ones for
$d^2 < 2/p$.

In our case, $d$ and $\Lambda$ are determined by the reduction Ansatz (\ref{Ansatz2}),
\be
d= \sqrt{\frac{2(p+n)}{pn}}, \hsp{2cm}
\Lambda = - n (n-1) \bK.
\label{values}
\ee
In order for $\Lambda$ to be negative, we are forced to take $\bK = +1$, {\it i.e.} the angular
part in (\ref{Ansatzcurved}) and (\ref{Ansatz2}) is the $n$-sphere $\S^n$. In this notation the
domain wall solution takes the form
\bea
&& \d s^2 \ = \ \Bigl[1 + \lambda \, z\Bigr]^2 \tg_{ij}\, \d x^i \d x^j
         + \d z^2,  \nnw
&& \e^\chi \ =  \ \Bigl[1 + \lambda \,
z\Bigr]^{-\sqrt{\frac{2pn}{p+n}}}, \label{AJS5bis} \eea with
$\lambda = p^{-1} \sqrt{(n-1)(n+p)}$. Note that the solutions  exist
for all values of $p$ between 1 and $D-4$ and that they all require
\emph{de Sitter} sliced domain walls\footnote{When we consider the
coordinate transformation $y=1+\lambda z$ the solution reads 
$\d s^2= \d y^2 + y^2\d \tilde s_{p+1}^2$ with $\d \tilde s_{p+1}^2$ being the metric
on $(p+1)$-dimensional de Sitter space. This can be confusing, since it is (almost) identical 
to flat space in Milne coordinates. However the reason
that these domain walls do not  describe flat space is because the
metric  $\d s^2= \d y^2 + y^2\d \tilde s_{p+1}^2$ only corresponds to flat space for a specific 
normalisation of the curvature of the de Sitter slice, namely $\tR_{p+1}=(p+1)p$. In all
our solutions the de Sitter curvature is in fact different.} with
worldvolume curvature \be \tLambda = \frac{1}{3} (n-1). \ee One can
understand the range for the values of $p$ as follows: $p$-branes
with $p<1$ are pointlike and cannot have a curved worldvolume, while
$p$-branes with $p> D-4$ have codimension 2 or smaller, such that
the angular part of the transverse space cannot be curved either.

Using the reduction relations (\ref{redrel}), we can write the
solution (\ref{AJS5bis}) as a purely gravitational solution in
$D=p+n+2$ dimensions,
\be 
\d \hs^2 = \Bigl(1+ \lambda\, z\Bigr)^{\frac{2p}{p+n}}\,
\tg_{ij}\, \d x^i \d x^j
             + \Bigl(1+ \lambda \, z \Bigr)^{-\frac{2n}{p+n}} \d z^2
             + \Bigl(1+ \lambda \, z\Bigr)^{\frac{2p}{p+n}} \d\Omega_n^2.
\ee
Applying  the coordinate transformation
\be
1+ \lambda\, z = r^\lambda,
\ee
the solution can be written in the more pleasing form
\be
\d \hs^2 =
r^{2\sqrt{\frac{n-1}{p+n}}}\, 
                        \tg_{ij}\, \d x^i \d x^j
\ + \
r^{2\sqrt{\frac{n-1}{p+n}}-2}\,
          \Bigr[\d r^2 + r^2 \d\Omega_n^2\Bigr],
\ee
where $\tg_{ij}$ satisfies the Einstein condition (\ref{4dimcurvature}).

The solution (\ref{AJS5bis}) is a so-called scaling solution, due to
the polynomial dependence in the scale factor $a(y)$. These solutions
are often found in FLRW cosmologies coupled to scalars in an
exponential potential. It is therefore no surprise that we find a
similar behaviour here, as the domain wall/cosmology correspondence
relates these two types of solutions \cite{SkT1, SkT2}. The fact that
we are dealing with scaling solutions will make it easy to
generalise them to the case with non-zero flux ($Q\neq 0$).

Recently an extension of this model has been considered in
\cite{Fonseca:2011ep} where the exponential potential contains a
linear combination of two scalar fields. A suitable rotation in
field space maps this model to the model with a single scalar field
in the exponential plus one free decoupled scalar field.

\sect{Curved branes with flux}

\subsection{The scaling solutions for general parameters}

A more interesting case is that of domain walls with two scalars in a double exponential potential,
as these can be interpreted as dilatonic $p$-branes with non-zero charge. Inspired by the scaling
solution of the previous section, we insist on both $\chi$ and $\phi$ depend logarithmically on
$a(z)$,
\begin{equation}
\phi = N_1 \log a(z), \qquad \chi = N_2\log a(z).
\end{equation}
Again the requirement that the potential scales
like $a^{-2}$ fixes the constants $N_1$ and $N_2$ as
\begin{equation}
N_1=\frac{2(c-d)}{bd}, \hsp{2cm}
N_2 = -\frac{2}{d},
\end{equation}
and the domain wall solution is then given by
\bea
&& \d s^2 \ = \ \Bigl[1 + \lambda \, z\Bigr]^2 \tg_{ij}\, \d x^i \d x^j
         + \d z^2,  \nnw
&& \e^\phi \ = \ \Bigl[1 + \lambda \, z\Bigr]^{\frac{2(c-d)}{bd}},
\hspace{1cm} \e^\chi \ =  \ \Bigl[1 + \lambda \,
z\Bigr]^{-\frac{2}{d}}, \label{DWQ} \eea provided that \bea &&
\lambda \ = \ bd\, \sqrt{\frac{-\Lambda}{2p\,(b^2 + c^2 -cd)}},\nn
[.3cm] && Q^2 \ = \ \frac{-2(c-d)d}{b^2 + c^2 -cd}\,  \Lambda, \nn
[.4cm] && \tLambda \ = \ \frac{b^2 + (c-d)^2 -\half pb^2 d^2
}{p^2(b^2 + c^2 -cd)}\, \Lambda.\label{conditions} \eea

In the reduction context, where $c$, $d$ and $\Lambda$ are given by (\ref{values2}) and
(\ref{values}), this solution takes the form
\bea
&& \d s^2 \ = \ \Bigl[1 + \lambda \, z\Bigr]^2 \tg_{ij}\, \d x^i \d x^j + \d z^2,  \nnw
&& \e^\phi \ = \ \Bigl[1 + \lambda \, z\Bigr]^{\frac{2p(n-1)}{(p+n)b}}, \hspace{1cm}
\e^\chi \ =  \ \Bigl[1 + \lambda \, z\Bigr]^{-\sqrt{\frac{2pn}{(p+n)}}},
\label{DWQ2}
\eea
where now
\bea
&& \lambda \ = \ \frac{b(p+n)}{p}\, \sqrt{\frac{(n-1)\,\bK}{(p+n)b^2 + 2(p+1)(n-1)}},\nn [.3cm]
&& Q^2 \ = \ \frac{4(n-1)^2 (p+n)}{(p+n)b^2 + 2(p+1)(n-1)}, \nn [.4cm]
&& \tLambda \ = \ \frac{n-1}{p}\,  \frac{ (p+n)b^2 - 2(n-1)^2 }{(p+n)b^2 + 2(p+1)(n-1)}\,\bK.
\label{parameters}
\eea
Again we find that necessarily the square root in the expression for $\lambda$ forces us to
take $\bK=+1$. We see therefore that in the case with flux, we find both positively and negatively
curved $p$-branes, depending on the value of the dilaton coupling $b$. In particular we can have 
de Sitter geometry if the dilaton coupling $b$ is big enough, $b>(n-1) \sqrt{2/(p+n)}$.

This solution can be lifted up to $D= p + n + 2$ dimensions as
\bea 
&& \d \hs^2 =r^{A} \, \tg_{ij}\, \d x^i \d x^j
      \ + \  r^{A-2}\,
              \Bigr[\d r^2 + r^2 \d\Omega_n^2\Bigr], \nnw
&& \e^\hphi =  r^{B},
\hsp{2cm} \hF_{a_1\dots a_n} = \frac{Q}{\sqrt{|\bh|}}\,
\bar \varepsilon_{a_1\dots a_n},
\eea
where $\tLambda$ and $Q$ are given
in (\ref{parameters}), $A$ and $B$ by
\be
A = \sqrt{\frac{4b^2(n-1)}{(p+n)b^2 + 2(p+1)(n-1)}}, \hsp{1.3cm}
B=\sqrt{\frac{4(n-1)^3}{(p+n)b^2 + 2(p+1)(n-1)}},
\ee
and the
worldvolume metric $\tg_{ij}$ satisfies the Einstein relation \bea
&& \tR_{ij} \, = \, -p \, \tLambda \, \tg_{ij}
           \hsp{2cm} \mbox{for} \ \ b^2> \tfrac{2(n-1)^2}{(p+n)}, \nnw
&& \tR_{ij} \, = \, + p \, \tLambda \,  \tg_{ij}
           \hsp{2cm} \mbox{for} \ \ b^2<\tfrac{2(n-1)^2}{(p+n)}.
\eea

\subsection{The scaling solutions in 10-dimensional supergravity}

The general solutions found in the previous subsection simplify
remarkably for the case of ten-dimensional Type IIA/B supergravity.
For the case of the RR fields, we have that
\be b= \frac{p-3}{2},
\ee
such that the conditions (\ref{conditions}) become
\bea
\lambda = \frac{p-3}{p}\sqrt{\frac{7-p}{2}}\,, \hsp{1cm}
Q^2 = (7-p)^2\,,\hsp{1cm}
\tLambda = -\frac{(7-p)(5-p)}{2p} \,.
\eea
These Einstein D-branes are negatively curved (AdS) for $p<5$ and positive
(dS) for $p>5$. The case $p=5$ is special, as it is (Ricci) flat. The solution is
given by
\bea && \d \hs^2 = r^{\frac{1}{2}} \, \tg_{ij}\, \d x^i \d x^j
      \ + \  r^{-\frac{3}{2}}\,
              \Bigr[\d r^2 + r^2 \d\Omega_3^2\Bigr], \nnw
&& \e^\hphi = r,
\hsp{2cm} \hF_{a_1\dots a_3} = \frac{2}{\sqrt{|\bh|}}\,
\bar \varepsilon_{a_1\dots a_3},
\eea
with $\tR_{ij}= 0$. It can easily be shown that this corresponds
to the near-horizon geometry of the standard (Ricci-flat) D5-brane. In section
\ref{beyond} we will show that this results is in fact quite general, in the sense
that the flat scaling solutions of the equations (\ref{EOM}) yield the near-horizon
geometries of the standard $p$-branes.

The D7 and D8 are also special, as the angular part of the transverse space is either
one- or zero-dimensional and can therefore not be curved. Nevertheless, curved scaling
solutions can still be constructed in these cases, by considering solutions with a single
exponential scalar potential, coming from the flux form field in higher dimensions ({\it i.e.}
turning off the exponential proportional to $\bK$). We will not give the exact expressions
here, as they can be easily derived form the general results of section \ref{Q=0}. Yet it is
useful to remark that the flat D8-brane solution is in fact a scaling
solution as well. This is not true for the other flat BPS branes, as
we explain in section \ref{beyond}.

Note that the metric becomes ill-defined for $p=3$. It is well known that the flat
D3-brane does not have a scaling solutions as its near-horizon limit due to the absence of the 
dilaton. It is easy to see that the same argument extends to the curved case, for which no scaling 
solutions will exist. The case $p=3$ should therefore be excluded
from our solutions, as it falls beyond the Ansatz used here.

Finally, the remaining F1 and the NS5-brane are easy to discuss: as they couple (magnetically)
to the NSNS 7- and 3-form respectively, we have that
\begin{align}
&\text{F}1: \qquad b = -1\,,\qquad p=1\,,\nonumber\\
&\text{NS}5 : \quad \,\,b= +1\,,\qquad p=5\,,
\end{align}
and their solutions are the same as the D1 and D5-brane, up to a minus sign in the dilaton
exponential (as can be expected from S-duality). We therefore find a negatively curved (AdS)
F1 and a Ricci-flat NS5-brane.


\sect{Timelike/spacelike $p$-brane correspondence}

So far we have only considered timelike $p$-branes, whose worldvolume is Lorentzian and which
have static geometries. Similarly there exist spacelike brane solutions \cite{Gutperle:2002ai},
which have Euclidean worldvolume and a time-dependent geometry ({\it i.e.} the warp factors in
the two-block Ansatz depends solely on time)
\bea && \d\hat s^2 \, =
\, \e^{2A(t)} \, \tg_{ij} (x)\,\d x^i \d x^j \,
           +  \, \e^{2B(t)} \Bigl( -\d t^2 + t^2 \d\Sigma_n^2\Bigr),
\label{AnsatzcurvedII}
\eea
It should be clear that the same procedure applied to timelike branes can equally well
be performed on the spacelike ones: we can dimensionally
reduce over the slicing $\d\Sigma_n^2$ of the transverse space, which will give rise to a
$(p+2)$-dimensional FLRW solution after reduction,
\bea
&& \d s^2 \, = - \, f^2(t) \d t^2 + a^2(t) \tg_{ij}(x) \,\d x^i \d x^j \, .
\label{FLRWAnsatz}
\eea
 These metrics are solutions to a set of differential equations analogous to (\ref{EOM}),
which differ from that latter in the sign of $\tilde{\Lambda}$ and the potential. Obviously,
the time coordinate $t$ is the analogue of the transverse direction $z$ in the domain wall
case, as it is transverse to the spatial sections of the FLRW metric.
Note that now $\tilde{\Lambda}$ has the usual interpretation of the FLRW spatial curvature,
usually denoted $k$.

\begin{figure}
\centering
\includegraphics[width=.7\textwidth]{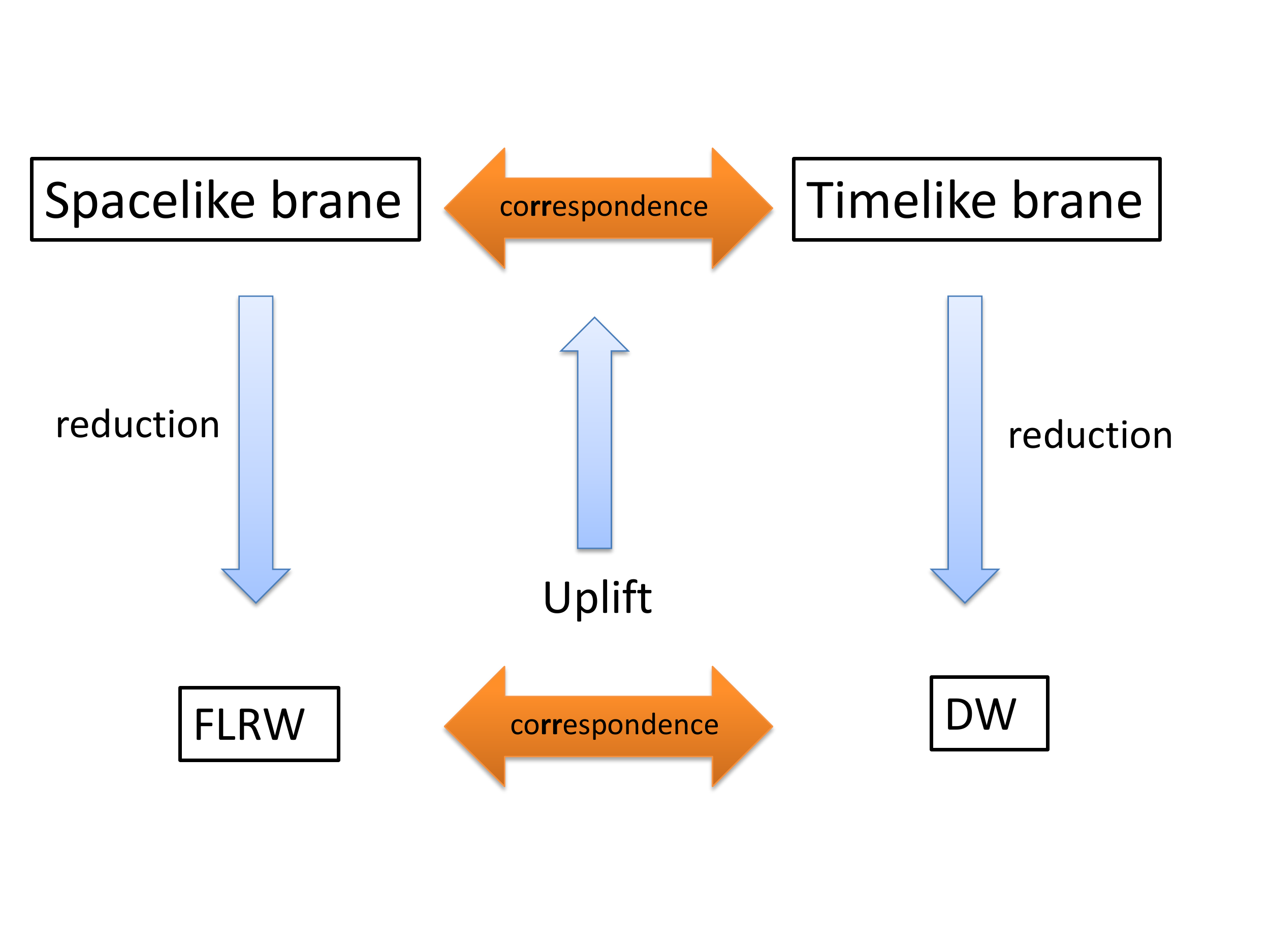}
\caption{\emph{The general timelike/spacelike brane correspondence
from the DW/FLRW correspondence.}}
\label{Figure1}
\end{figure}

This analogy is part of what has been named the domain
wall/cosmology correspondence \cite{SkT1,SkT2}. Interestingly, this
correspondence here gets uplifted (and hence generalised) to a more
general \emph{timelike/spacelike $p$-brane}
correspondence,\footnote{This has been discussed briefly before in
\cite{Janssen:2007rc}.} as is schematically depicted in Figure
\ref{Figure1}.

Let us quickly discuss this for the simplest case, with $Q^2=0$. The
DW/FLRW correspondence tells us that domain wall scaling
solutions get mapped to FLRW solutions with the opposite sign of
the potential and opposite sign of the worldvolume curvature ({\it i.e.} wall curvature
for domain walls and spatial curvature of FLRW metrics). This implies that for
$p=1,\ldots, 6$ we have a set of S-brane scaling solutions with opposite spatial
curvature $\tilde{\Lambda}$ (negative in the $Q^2 = 0$ case) and a \emph{negatively curved
slicing} in the space transverse to the brane, $\bK<0$.
The scale factor takes the form
\begin{equation}
a(t)=1+ \lambda t \,,
\end{equation}
with the appropriate sign flips for $\bK$ and $\tLambda$ in the expression for $\lambda$.

In the case
with $Q^2 \neq 0$, we cannot reverse the sign of the flux contribution.
However if we would nonetheless insist on doing so, this would
bring us to supergravity theories with ghostlike kinetic terms.
In this context the DW/COSM correspondence has been discussed as
well \cite{Skenderis:2007sm, Bergshoeff:2007cg}.

Having hyperbolically sliced transverse directions, $\bK<0$, is
often -though not always- part of the definition of an S-brane, as
flat S-branes  asymptote to flat
space in Milne coordinates at late times. This is  similar in the
way that timelike $p$-branes with $\bK>0$
approach flat space in radial coordinates at spatial infinity.

\sect{Scaling solutions as near-horizon regions}
\label{beyond}

The above timelike/spacelike brane correspondence allows us to understand better 
the scaling solutions constructed in this paper, using knowledge from the cosmology side. 
For the case of FLRW
cosmologies with exponential potentials, it is known that scaling
solutions do not describe all solutions but they are critical points
of an autonomous system of equations of motion, such that they
correspond to the late-time or early time behavior of the most
general solution. In other words, they can describe attractors or
repellers. We refer to \cite{Collinucci:2004iw, Hartong:2006rt} for
a treatment of this in a  general setting that include the scalar
potentials discussed here.

In the appendix we have written down the autonomous system 
corresponding to the equations of motion (\ref{EOM})
and derived its critical points.  In the curved case we reproduce exactly
the scaling solutions $a(z)\sim z$, presented in the previous sections.
In the flat case ($\tLambda=0$) we can also find scaling
solutions, that now go like 
\begin{equation}
a(z)\sim z^{\ell}\qquad \text{with}\qquad
\ell=\frac{(9-p)}{(p-3)^2}\,.
\end{equation}
Notice that for the 5-brane we have that $\ell = 1$, which agrees nicely with the fact that 
the same solutions appears also as the Ricci-flat case of the Einstein branes (which all have 
$\ell = 1$). In a certain sense the 5-branes are the intersection of the two classes of solutions. 

In order to interpret these flat scaling solutions, let us uplift them to 10 dimensions. 
Plugging the
expressions for $a(z)$ and $\chi(z)$ in the full 10-dimensional
metric and performing the following coordinate
transformation
\begin{equation}
z =  r^{\frac{(p-3)^2}{2p}}\,,
\end{equation}
we find the near-horizon solutions for the standard extremal D$p$-branes:
\begin{equation}
\d s^2= H^{\frac{p-7}{8}}\d s_{p+1}^2 + H^{\frac{p+1}{8}}\Bigl(\d
r^2 + r^2\d\Omega^2\Bigr)
\end{equation}
with
\begin{equation}
H\sim r^{p-7}\,.
\end{equation}
The full extremal brane solutions are given by $H=\alpha +\beta
r^{p-7}$, but only the near-horizon case $\alpha=0$ corresponds to a
scaling solution (or equivalently, a critical point of the
autonomous system).

There are two important lessons that we can draw from this:
\begin{enumerate}

\item The scaling solutions (both flat as curved ones) are not the general solutions, but 
describe the near-horizon of these. We just showed this explicitly for the flat case, but there
is no reason to assume that it would be different for the curved case. In fact the autonomous
system formalism tells us so: the critical points of the flow equations (the scaling solutions) 
are particular limits of the full solutions, which in turn are described by the flow 
lines of the autonomous system. In other words, there must exist extensions of our curved Einstein 
branes that interpolate between different scaling solutions for small and large $r$. 

\item In those cases we could not obtain a curved
$p$-brane scaling solution ({\it e.g.} AdS curved branes without
flux, or the $p=3$ case with flux), there still exist curved brane solutions: they just do not 
not have a scaling regimes. To understand the full space of solutions for
either sign of the curvature we refer to \cite{ST} for a treatment
of the single exponential case.
\end{enumerate}
At the moment we have not been able to find the full analytic
solution interpolating between scaling solutions but hope to report
on it in the future.


\sect{Discussion}

In this paper we have constructed many new solutions that have the
interpretation of $p$-brane whose worldvolume is a curved Einstein
space. The exact solutions we were able to find correspond to
scaling solutions. This means that, when the brane is reduced to a
(curved) domain wall, the warp factor $a(z)$ is a simple power-law
$z^{\ell}$,
\begin{equation}
\d s^2 = \d z^2 + z^{\ell} \tg_{ij}\d x^i\d x^j\,,
\end{equation}
where the $(p+1)$-dimensional wall geometry is Einstein
\begin{equation}
\tR_{ij} = p\, \tilde{\Lambda}\, \tg_{ij}\,.
\end{equation}
When the curvature is non-zero the power-law necessarily has
$\ell=1$.

Scaling solutions uplift to $p$-brane near-horizons. This is explicitly
confirmed in section 6 for the known case of the standard flat
branes, but it is natural to extrapolate this to the case of curved
branes. A formal proof of this was achieved by rewriting the
equations as an autonomous system for which the scaling solutions
are the attractor critical points. Therefore there exist more
general solutions that interpolate between a proper scaling solution
(the near-horizon) and a non-proper scaling solution, with infinite
fields (spatial infinity).

Our approach can be generalised trivially to S-branes. Whereas
timelike branes reduce to domain walls, spacelike branes reduce to
FLRW cosmologies. The domain wall/cosmology correspondence in this
framework gets generalised to a general timelike/spacelike brane
correspondence. In this context curved branes are perhaps more
natural since they correspond to the curvature of the spatial part
of the FLRW metric.

Amongst our explicit curved timelike brane solutions we have many
that have de Sitter worldvolumes. This was the case for all fluxless
brane solutions and for the solutions with flux when $p=6$. However,
when we consider an internal slicing that is not spherical,
$\bK \neq 1$, we can get de Sitter curved
worldvolumes. De Sitter branes have appeared earlier in
\cite{AJS} and \cite{Neup1}-\cite{Minamitsuji:2011gp} however the solutions in
\cite{Neup1, Neup2} are written in more complicated coordinates. Our
approach makes clear that these solutions are not the most general,
but should be seen as near-horizons. These de Sitter solutions can
not be regarded as warped de Sitter compactifications since the
space transverse to the brane worldvolume is necessarily
non-compact. However, as warped non-compactifications, such a
solution might be relevant if gravity is localised sufficiently and
there is some understanding of the presence of gauge forces living
on these branes.

Finally we want to compare our method with another well known method
that employs dimensional reduction. This method is based on reducing
flat $p$-branes over their worldvolume \cite{Bergshoeff:2008be} (see
also \cite{Janssen:2007rc}), a technique inspired from the special
case of black holes \cite{Breitenlohner:1987dg}. Since the
worldvolume is flat, this does not generate a scalar potential,
rather one just obtains a sigma model that is solvable and whose
integrability can be understood in a formal way using group theory
and the Hamilton-Jacobi formalism \cite{Chemissany:2010zp}. This
works for a very large class of generalisations with much less
worldvolume symmetries \cite{Bergshoeff:2008be}. This is in contrast
with the technique used in this paper, where we reduce over the
curved slice in the transverse space instead of the worldvolume
(see for a discussion of the two approaches
\cite{Bergshoeff:2008zza}) . However we could equally reduce over
the curved worldvolume here and it would generate a scalar potential
as well. It would be interesting to see whether in this case the
equations of motion could be fully solvable as well.


\vspace{1cm}
\noindent
{\bf Acknowledgements}\\
We wish to thank Admiral Freebee for inspiration.
 The work of W.C. is supported in part by the FWO -
Vlaanderen, Project No. G.0651.11, and in part by the Federal Office
for Scientific, Technical and Cultural Affairs through the
Interuniversity Attraction Poles Programme - Belgian Science Policy
P6/11-P.
The work of B.J.
is partially supported by the M.E.C. under contract FIS2010-17395
and by the Junta de Andaluc\'{\i}a groups P07-FQM 03048 and
FQM-6552.

\appendix
\renewcommand{\theequation}{\Alph{section}.\arabic{equation}}
\sect{Autonomous equations of motion}

The autonomous variables are given by
\begin{align}
& X_1=\frac{\phi'}{H\sqrt{2p(p+1)}}\,,\qquad X_2=\frac{\chi'}{H\sqrt{2p(p+1)}}\,,\nonumber\\
& Y_1=\tfrac{1}{2}\frac{Q^2\e^{b\phi+c\chi}}{p(p+1)H^2}\,,\qquad
Y_2=\frac{-\bK n(n-1)\e^{d\chi}}{p(p+1)H^2}\,,
\end{align}
where $H=a'/a$. One can check that the  equations become first-order
equations
\begin{align}
 \dot{X}_1=& X_1(p+1)\Bigl(X_1^2 +X_2^2 -1 +\frac{\epsilon\tilde{\Lambda}}{(p+1)\dot{a}^2} \Bigr)
    + b\epsilon\sqrt{\tfrac{p(p+1)}{2}} Y_1 \,,\nonumber\\
 \dot{X}_2=& X_2(p+1)\Bigl(X_1^2 +X_2^2 -1 +\frac{\epsilon\tilde{\Lambda}}{(p+1)\dot{a}^2} \Bigr)
       + c\epsilon\sqrt{\tfrac{p(p+1)}{2}} Y_1 + d\epsilon\sqrt{\tfrac{p(p+1)}{2}} Y_2\,,\nonumber\\
 \dot{Y}_1=& \sqrt{2p(p+1)}\,Y_1\Bigl(bX_1 + cX_2\Bigr)  + 2(p+1)Y_1\Bigl(X_1^2 +X_2^2
       +\frac{\epsilon\tilde{\Lambda}}{(p+1)\dot{a}^2}\Bigr)\,,\nonumber\\
 \dot{Y}_2=& d\sqrt{2p(p+1)}\,\, Y_2X_2 + 2(p+1)Y_2\Bigl(X_1^2 +X_2^2
     +\frac{\epsilon\tilde{\Lambda}}{(p+1)\dot{a}^2}\Bigr)\,,\label{autonomous4}
\end{align}
together with the constraint
\begin{equation}\label{constraint}
X_1^2 + X_2^2-\epsilon Y_1 -\epsilon Y_2
+\epsilon\frac{\tilde{\Lambda}}{\dot{a}^2} = 1\,.
\end{equation}
We used a dot to denote differentiation with respect to $\ln(a)$
whereas a prime we have been using to denote differentiation with
respect to $z$. We furthermore introduced $\epsilon$, which takes
value $\epsilon=+1$ for domain walls and $\epsilon=-1$ for FLRW
cosmologies.

The constraint equation (\ref{constraint}) is not really an
independent equation. One can show that, when the initial conditions
obey the constraint, so will the evolution. This can be proven by
taking the derivative of the constraint equation and showing that it
is automatically satisfied using the equations (\ref{autonomous4}).

When $\tilde{\Lambda}=0$ we have a true autonomous system whose
dynamics can be understood partially from the critical points,
defined as solutions with constant $X,Y$. These are the simplest
solutions and general solutions interpolate between these critical
points. These critical points come in two kinds: those with finite
values of the scalars and those with infinite valued scalars
\cite{Collinucci:2004iw}. The latter class can describe solutions at
spacelike or timelike infinity. Let us here discuss the first class
of critical points, with finite scalars. Solving the algebraic
equations that one gets when
$\dot{X}_1=\dot{X}_2=\dot{Y}_1=\dot{Y}_2=0$ gives the following
solution
\begin{align}
&Y_1 =\epsilon \frac{2-2(p+1)\ell}{p(p+1)\ell^2}\Bigl(b^{-2}-cb^{-2}d^{-1}\Bigr)\,,\nonumber\\
&Y_2 =\epsilon \frac{2-2(p+1)\ell}{p(p+1)\ell^2}\Bigl(-b^{-2}cd^{-1}+c^2b^{-2}d^{-2}+d^{-2}\Bigr)\,,
\nonumber\\
&X_1 =-\frac{1}{\ell}\sqrt{\tfrac{2}{p(p+1)}}\Bigl(b^{-1}-cb^{-1}d^{-1}\Bigr)\,,\nonumber\\
&X_2 =-\frac{1}{\ell}\sqrt{\tfrac{2}{p(p+1)}}\,\,d^{-1}\,, \nonumber\\
&\ell=\frac{2}{p}\Bigl(b^{-2}+d^{-2}+b^{-2}c^2d^{-2}-2b^{-2}cd^{-1}\Bigr)\label{XYexpressions}\,.
\end{align}
When we consider the following equation
\begin{equation}
\frac{H'}{H^2}=-\epsilon \frac{\tilde{\Lambda}}{{a'}^2}-(p+1)X_1^2
-(p+1)X_2^2\,.
\end{equation}
for the case  $\tilde{\Lambda}=0$ we find that a critical point must
obey
\begin{equation}
\frac{H'}{H^2}\equiv-\ell^{-1}=\text{constant}\quad \Longrightarrow
\quad a(z)\sim z^{\ell}\,.
\end{equation}
So the scale factor $a$ is given by a simple power-law. These
solutions are called scaling solutions since every term in the
action, or equations of motion, scales in the same way. The scaling
solutions, in terms of the  fields, read
\begin{equation}
\phi(z)=X_1\ell\sqrt{p(p+1)}\,\,\ln(z) +\phi(1)\,, \qquad
\chi(z)=X_2\ell\sqrt{p(p+1)}\,\,\ln(z) +\chi(1)\,.
\end{equation}
Let us now consider the case of standard flat timelike $p$-branes
$\tilde{\Lambda}=0, \bK=1$. Then we find that the values for the
$Y$'s are such that a solution exists for $p=0,\ldots 6$ excluding
$p=3$. If we plug in the specific values for $b,c$ and $d$ we find
that
\begin{equation}
\ell=\frac{(9-p)}{(p-3)^2}\,.
\end{equation}

The same strategy still applies when $\tilde{\Lambda}\neq 0$. For
the latter case we necessarily have that (see
e.g.~(\ref{constraint})) that $\tilde{\Lambda}\dot{a}^{-2}$ is
constant. Therefore
\begin{equation}
a(z)\sim z\,.
\end{equation}
One can easily check, using the techniques of \cite{Hartong:2006rt}
that we obtain exactly the scaling solutions given in the previous
sections. All expressions for $X$ and $Y$ (\ref{XYexpressions}) are
still valid if we use $\ell=1$.

\end{document}